\documentclass[aps,prb,twocolumn,showpacs,superscriptaddress,floatfix]{revtex4-1}
\usepackage{graphicx,graphics}
\usepackage{dcolumn}
\usepackage{amsmath,amssymb,amsfonts}
\usepackage{latexsym,verbatim}
\usepackage{bm}
\usepackage{color}
\usepackage{ulem}
\usepackage[breaklinks=true,colorlinks,citecolor=blue,linkcolor=blue,urlcolor=blue]{hyperref}

\def\be{\begin{eqnarray}}
\def\ee{\end{eqnarray}}

\begin{document}
\title{Stripe glasses in ferromagnetic thin films}
\author{Alessandro Principi}
\email{aprincipi@science.ru.nl}
\affiliation{Institute for Molecules and Materials, Radboud University, Heijndaalseweg 135, 6525 AJ, Nijmegen, The Netherlands}
\author{Mikhail I. Katsnelson}
\affiliation{Institute for Molecules and Materials, Radboud University, Heijndaalseweg 135, 6525 AJ, Nijmegen, The Netherlands}
\begin{abstract}
Domain walls in magnetic multilayered systems can exhibit a very complex and fascinating behavior.
For example, the magnetization of thin films of hard magnetic materials is in general perpendicular to the thin-film plane, thanks to the strong out-of-plane anisotropy, but its direction changes periodically, forming an alternating spin-up and spin-down stripe pattern. The latter is stabilized by the competition between the ferromagnetic coupling and dipole-dipole interactions, and disappears when a moderate in-plane magnetic field is applied. It has been suggested that such a behavior may be understood in terms of a self-induced stripe glassiness. In this paper we show that such a scenario is compatible with the experimental findings.
The strong out-of-plane magnetic anisotropy of the film is found to be beneficial for the formation of both the stripe-ordered and glassy phases. 
At zero magnetic field the system can form a glass only in a narrow interval of fairly large temperatures. An in-plane magnetic field, however, shifts the glass transition towards lower temperatures, therefore enabling it at or below room temperature.
In good qualitative agreement with the experimental findings, we show that a moderate in-plane magnetic field of the order of $30~{\rm mT}$ can lead to the formation of defects in the stripe pattern, which sets the onset of the glass transition.
\end{abstract}
\maketitle

\section{Introduction}
%
%In its common use, the word ``frustration'' has generally a negative connotation. It encodes the ensemble of feelings experienced by somebody who is forbidden by some external constraint to achieve his goal and forced to choose between many non-optimal solutions.
In its common use, the word ``frustration'' has generally a negative connotation. It encodes the ensemble of feelings experienced by somebody who, subject to various constraints that cannot be simultaneously satisfied, is forced to choose one of many non-optimal solutions. 
Physical systems can also experience something similar, although in a much less dramatic way.
%A similar situation is realized in physical systems subject to competing constraints that cannot be satisfied altogether.
The classical examples are Ising spin glasses.~\cite{Mezard_Parisi_Virasoro_book} In the standard formulation, the frustration is introduced by the random signs of couplings between the local magnetic moments. Indeed, each spin interacts ferromagnetically with a (randomly chosen) set of its neighbors, and antiferromagnetically with the others. Therefore, for a generic choice of the couplings, it is not clear in which direction it should orient in order to lower its energy. This is a ``cooperative problem'': to find the ground state energy it is indeed necessary to consider the system as a whole, and not at the level of the single spin. This problem is extremely complicated, since the system exhibits a large number of configurations energetically very close to the ground state. It is well known that a glass forms in these conditions.~\cite{Mezard_Parisi_Virasoro_book} Below a characteristic temperature $T_{\rm A}$ the system is indeed stuck in each of the exponentially many {\it metastable} configurations for a characteristic time $\tau_{\rm w}$, and it eventually freezes in one of them below the Kauzmann temperature~\cite{Kauzmann_chemrev_1948} $T_{\rm K}$, when $\tau_{\rm w}\to \infty$.

%In its common use, the word ``frustration'' entails a bad connotation. In physics, instead, the frustration experienced by a system, for example subject to competing interactions, can make a new and very intriguing behavior emerge. This is the case of, e.g., Ising spin glasses.~\cite{Mezard_Parisi_Virasoro_book} In the standard formulation, the frustration of a Ising system arises from the random sign of the couplings between local magnetic moments. Each of the spins interacts ferromagnetically with some of its neighbors, and antiferromagnetically with the others. Therefore, for a general choice of the couplings, spins have no preferential direction, and are randomly distributed. The large number of almost-equivalent configurations, separated in phase space by large barriers, leads to the emergence of a glassy behavior. Below a characteristic temperature $T_{\rm A}$ the system is indeed frozen in one of the exponentially many {\it metastable} configurations for a characteristic time $\tau_{\rm w}$, and eventually freezes in one of them, thus forming a static glass, below the Kauzmann temperature $T_{\rm K}$.

Randomizing the couplings is not the only way to introduce frustration in a system. Spins ordered in perfect lattices can also exhibit a glassy behavior due to the geometric frustration they experience if the signs of the couplings between them are properly chosen (like, e.g., in the antiferromagnetic triangular net~\cite{Wannier_pr_1950}). Another way to introduce frustration is to have competing interactions on different length scales. This is the case, e.g., of
%, a situations which is naturally realized in, e.g., multilayered thin films of alternating ferromagnetic layers,
spins locally coupled by a ferromagnetic exchange but subject to a long-range dipole-dipole interaction.~\cite{Choe_prb_1999,Ng_prb_1995,Whitehead_jpcm_1994} The minimization of the energy requires the total magnetization to be zero, a condition that prevents the formation of a uniformly polarized state. A striped phase is realized at low temperature: spin-up stripes alternate with spin-down ones.~\cite{Kittel_pr_1946,Garel_prb_1982,Yafet_prb_1988,Kaplan_jmmm_1993}
%, while the thickness of the domain walls is due to the energetics. 
The modulus of the wavevector of the stripes is determined by the competition between the local and long-range interactions, while its direction is due to the spontaneous breaking of the rotational symmetry.

%Spins antiferromagnetically coupled, and ordered in triangular lattices, experience a geometric frustration that can in principle lead to a glassy behavior. A more general situation is that of competing interactions on different length scales, which has attracted a large deal of interest in the last decades. This is the case, for example, of spins coupled ferromagnetically and subject to a long-range dipole-dipole interaction. The latter tends to introduce an antiferromagnetic order on larger scales, determined by the microscopic parameters of the model. Such a system is believed to coalesce in a striped phase at low temperature. Spin-up stripes alternate with spin-down ones, and the thickness of the domain walls is due to the energetics. The modulus of the wavevector of the stripes is determined by the competition between the local and long-range interactions, while its direction is due to the spontaneous breaking of the rotational symmetry.

The frustration due to the two competing interactions at different length scales gives rise,~\cite{Monasson_prl_1995,Schmalian_prl_2000,Westfahl_prb_2001,Westfahl_prb_2003,Wu_prb_2004} in the thermodynamic limit, to an exponentially large number of local minima in the phase space. Such states are populated according to the Boltzmann distribution. At sufficiently large temperature, the large number of states compensates for their little statistical weight and allows the striped phase to ``melt'', thus forming a stripe glass. This state can be characterized by the entropy lost by the system ($S_{\rm c}$) by finding itself in a local (but not global) minimum of the potential landscape. $S_{\rm c}$ is also named ``configuration entropy''. 

%The configurational entropy can be determined, at the mean-field level, with the replica approach which was first introduced in Ref.~\onlinecite{Monasson_prl_1995}, and was successfully used in Refs.~\onlinecite{Schmalian_prl_2000,Westfahl_prb_2001,Westfahl_prb_2003,Wu_prb_2004}. The introduction of $n$ replicas allows, in the limit $n\to 1$, to account for the statistical weight (and therefore for the entropy contribution) of each metastable state.

The competition between interactions at different length scales arises naturally in multilayered thin films of alternating ferromagnetic layers.~\cite{Choe_prb_1999,Ng_prb_1995,Whitehead_jpcm_1994,Kittel_pr_1946,Garel_prb_1982,Yafet_prb_1988,Kaplan_jmmm_1993} These systems have attracted a huge deal of interest in the recent past:~\cite{Seul_science_1995,Donzelli_jap_2003,Hellwig_physB_2003,Favieres_jap_2002,Onoue_jap_2002,Belliard_jap_1997,Barnes_jap_1994,Hamada_jmmm_1999,Ausanio_jmmm_2001,Zeper_jap_1989,Louail_jmmm_1997} many experimental techniques allow a real-time probing of the domain wall structure,~\cite{Ploessl_jap_1993,Tonomura_prb_1982,Scheinfein_prb_1991,Schafer_IEEE_1992,Eimuller_jap_2000,Castrucci_prb_2002,Asenjo_prb_2000,Hamada_jmmm_2002,Ausanio_jmmm_2002,Asenjo_jmmm_1999}
 thereby enabling the study of the domain-wall dynamics and evolution. From a theoretical standpoint, the equilibrium state of these heterostructures is completely determined by the energetics: striped domain-wall structures can form in layered thin films because of the complex interplay between magnetostatic energy, domain-wall energy, magnetocrystalline anisotropy and dipole-dipole interaction.~\cite{Brucas_prb_2004,Brucas_prb_2008,Hafermann_apl_2009,Prudkovskii_epl_2006} The competition between local ferromagnetic and long-range dipole-dipole interactions leads to the formation of striped domain walls. A chaotization of the domain-wall pattern is observed when a sufficiently strong in-plane magnetic field $H_\parallel$ is applied, with defects starting to appear at a field as low as $H_\parallel \sim 30~{\rm mT}$.~\cite{Brucas_prb_2004,Seul_prl_1992,Kashuba_prb_1993} 
Following the general idea of self-induced stripe glassiness,~\cite{Schmalian_prl_2000} it has been speculated~\cite{Brucas_prb_2004} that such behavior represents the onset of a glass transition. In this paper we show that this scenario is actually compatible with the experimental observations. We find that (i) the perpendicular magnetocrystalline anisotropy, which can be very strong in layered thin films, plays a key role and is beneficial for the formation of a stripe glass, and (ii) that the in-plane magnetic field can be used to trigger the glass transition.

The paper is organized as follows: in Sect.~\ref{sect:model} we introduce a minimal model that describes a magnetic system subject to competing interactions. An additional constraint forces the local magnetization to be a constant, and introduces (taking into account the Gaussian fluctuations around the mean-field value) a fourth-order coupling between the components of the magnetization. By means of the replica trick~\cite{Mezard_Parisi_Virasoro_book} we derive the expression of the configurational entropy.~\cite{Schmalian_prl_2000} The fourth-order coupling is fundamental to break the replica symmetry and obtain a finite $S_{\rm c}$. 
The latter is evaluated in Sect.~\ref{sect:results}, where we also show the phase diagram of the model. We find that the in-plane magnetic field can trigger the glass transition by shifting it to lower temperatures. When the transition temperature equals the room temperature, defects in the domain-wall pattern appear and a glass forms. The value of the critical field is very close to that applied experimentally, and for which the chaotization of the pattern is observed. Finally, Sect.~\ref{sect:conclusions} summarizes the main findings of this work.
 
% The same frustration can also lead to the formation of a striped glass phase. Moreover, layered thin films allow to tune the magnetocrystalline anisotropy. Some of these heterostructures show indeed a strong perpendicular anisotropy which, as we will show, is beneficial for the formation of a stripe glass.

\section{The model}
\label{sect:model}
In this section we introduce a minimal continuum model of a two-dimensional (2D) magnetic system. At the microscopic level, we assume the spins to be locally coupled ferromagnetically (at the nearest-neighbor level) and subject to a long-range dipole-dipole interaction. Going to the continuum limit, we introduce the position-dependent magnetization field, which results from the average over many microscopic spins and on which the Hamiltonian depends.~\cite{Prudkovskii_epl_2006} We assume the magnetization to be nearly constant and only its direction, denoted with the unit vector ${\bm m}({\bm r})$, to change. The partition function is given by $Z = \int {\cal D}{\bm m} \int {\cal D}\lambda \exp\{-\beta {\cal H}[{\bm m},\lambda]/2\}$, where $\beta = (k_{\rm B} T)^{-1}$ is the inverse temperature ($k_{\rm B}$ is the Boltzmann constant) and the Hamiltonian is~\cite{Prudkovskii_epl_2006}
\begin{eqnarray}
&&
{\cal H}[{\bm m},\lambda] \! = \!
\int \!\! d{\bm r} \Big\{ J \big[ \partial_i m_j({\bm r}) \big] ^2 - K m_z^2({\bm r}) - 2 {\bm h}({\bm r}) \cdot {\bm m}({\bm r}) \Big\}
\nonumber\\
&&
+ \frac{Q}{2\pi} \int \!\! d{\bm r} d{\bm r}' m_z({\bm r}) \left[ \frac{1}{|{\bm r}-{\bm r}'|} - \frac{1}{\sqrt{d^2+|{\bm r}-{\bm r}'|^2}} \right] m_z({\bm r}')
\nonumber\\
&&
+ \int d{\bm r} \Big\{ \lambda({\bm r}) \big[ {\bm m}^2({\bm r}) - 1 \big] \Big\}
~.
\end{eqnarray}
In momentum space space it reads
\begin{eqnarray} \label{eq:3D_action_def}
{\cal H}[{\bm m},\lambda] &=&
%\sum_{{\bm q},{\bm q}'} m^i_{\bm q} \Bigg\{ \Big[J q^2 \delta_{ij} - \left(K  - Q \frac{1-e^{-qd}}{q}\right) \delta_{i,z} \delta_{j,z} \Big]
%\nonumber\\
%&&
%\times
%\delta_{{\bm q},{\bm q}'} + \lambda_{{\bm q}-{\bm q}'} \delta_{ij} \Bigg\} m^j_{-{\bm q}'}  - \sum_{\bm q} {\bm h}_{-{\bm q}} \cdot {\bm m}_{\bm q} - \sum_{\bm q} \lambda_{\bm q}
%\sum_{{\bm q},{\bm q}'} \sum_{i,j} m^i_{\bm q} \Big\{ \big[G^{(0)}(q)\big]^{-1}_{ij} \delta_{{\bm q},{\bm q}'} + \lambda_{{\bm q}-{\bm q}'} \delta_{ij} \Big\} m^j_{-{\bm q}'}  
\sum_{{\bm q},{\bm q}'} \sum_{i,j} m^i_{\bm q} \big[G^{(0)}({\bm q},{\bm q}', \lambda)\big]^{-1}_{ij} m^j_{-{\bm q}'}  
\nonumber\\
&-&
\sum_{\bm q} (2{\bm h}_{-{\bm q}} \cdot {\bm m}_{\bm q} + \lambda_{\bm q})
~,
\end{eqnarray}
where 
\begin{eqnarray}
\big[G^{(0)}({\bm q},{\bm q}', \lambda)\big]^{-1}_{ij}&=& \left[J q^2 \delta_{ij} - \left(K  - Q \frac{1-e^{-qd}}{q}\right) A_{ij} \right]
\nonumber\\
&\times&
\delta_{{\bm q},{\bm q}'}
+ \lambda_{{\bm q}-{\bm q}'} \delta_{ij} 
~.
\end{eqnarray}
Here $J>0$ is the ferromagnetic coupling, $K$ is the out-of-plane anisotropy, $Q$ characterizes the strength of the dipole-dipole interaction, $d$ is the film thickness, and $A_{ij} = \delta_{i,z} \delta_{j,z}$ is the anisotropy matrix. Finally, ${\bm h}({\bm r})$ is the local magnetic field. The ``slave field'' $\lambda({\bm r})$ is introduce to constrain the magnetization to satisfy the equality $|{\bm m}({\bm r})| = 1$. 
%Note that the ferromagnetic Ising model in the presence of the dipole-dipole interaction can be obtained from Eq.~(\ref{eq:3D_action_def}) by assuming ${\bm m}({\bm r})$ to be a scalar. In this case the term proportional to $K$ is just a constant and can be forgotten.

Following Ref.~\onlinecite{Schmalian_prl_2000}, the configurational entropy is defined as the logarithm of the number of local (metastable) minima of the action. This quantity can be computed by introducing the pinning field ${\bm \psi}({\bm r})$, such that $Z[{\bm \psi}] = \int {\cal D}{\bm m} \int {\cal D}\lambda \exp\{-\beta {\cal H}_{\bm \psi}[{\bm m},\lambda]/2\}$, where $\beta = (k_{\rm B} T)^{-1}$ and
\begin{eqnarray}
{\cal H}_{\bm \psi}[{\bm m},\lambda] = {\cal H}[{\bm m},\lambda] + g \int d{\bm r} \big[ {\bm m}({\bm r}) - {\bm \psi}({\bm r}) \big]^2
~.
\end{eqnarray}
Here $g\to 0^+$ is a coupling constant which is sent to zero after the thermodynamic limit is taken. The free energy $F[{\bm \psi}] = -\beta^{-1} \ln Z[{\bm \psi}]$ is small if ${\bm \psi}({\bm r})$ is equal to a configuration of ${\bm m}({\bm r})$ which is also a minimum of the Hamiltonian ${\cal H}[{\bm m},\lambda]$. 

\begin{figure}[t]
\begin{center}
\begin{tabular}{c}
\includegraphics[width=0.99\columnwidth]{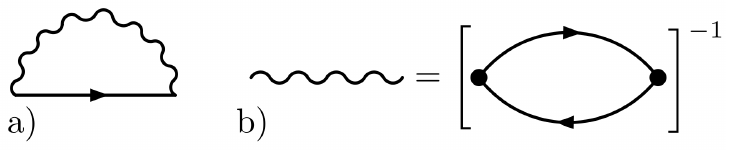}
\end{tabular}
\end{center}
\caption{
Panel a) the self-energy ${\tilde \Sigma}_{ij,\alpha\beta}(q)$ in the self-consistent screening approximation. The solid oriented line is the full Green's function ${\tilde G}_{ij,\alpha\beta}(q)$, while the wavy line represents the screened interaction ${\tilde D}_{\alpha\beta}(q)$.
Panel b) the screened interaction as the inverse of the polarizability ${\tilde \Pi}_{\alpha\beta}(q)$. The equations of panels a) and b) are solved self-consistently until convergence is reached.
\label{fig:one}}
\end{figure}

Scanning all the possible configurations of ${\bm \psi}({\bm r})$ we can gain information about the potential landscape of the magnetization, and in particular on the number of local minima of such landscape. The total free energy can be written as an average of $F[{\bm \psi}]$ weighted with the probability $Z[{\bm \psi}]$, {\it i.e.}
\begin{eqnarray} \label{eq:free_energy_average}
F_g = \frac{\int {\cal D} {\bm \psi} Z[{\bm \psi}] F[{\bm \psi}]}{\int {\cal D} {\bm \psi} Z[{\bm \psi}]}
~.
\end{eqnarray}
Note that, if the limit $g\to 0$ is taken {\it before} the thermodynamic limit, $F_{\rm eq} = \lim_{N\to \infty} \lim_{g\to 0} F_g$ is the free energy of the equilibrium state. On the contrary, if the two limits are interchanged, $F = \lim_{g\to 0} \lim_{N\to \infty} F_g$ contains the information about all the metastable states ({\it i.e.} the local minima of the potential landscape). It is important to note that the two limits coincide if the number of local minima does {\it not} grow exponentially with the volume of the system $V$.~\cite{Schmalian_prl_2000} Conversely, when the number of metastable states scales as $\sim e^V$, interchanging the two limits leads to two different results. The large number of states, indeed, compensates for the fact that their statistical weight is smaller than that of the equilibrium state. The difference between the two free energies is the configurational entropy, {\it i.e.}
%DEFINE N BEFORE
%
\begin{eqnarray}
S_{\rm c} = \beta (F - F_{\rm eq})
~,
\end{eqnarray}
which corresponds to the entropy lost by freezing the system in one of the exponentially many metastable states. Therefore, $S_{\rm c} \neq 0$ hints to a possible glassy phase. We now introduce
\begin{eqnarray} \label{eq:F_n}
F_n = - (\beta n)^{-1} \lim_{g\to 0} \ln \int {\cal D}{\bm \psi} Z^n[{\bm \psi}]
~,
\end{eqnarray}
where the function $Z[{\bm \psi}]$ in Eq.~(\ref{eq:F_n}) is replicated $n$ times. Therefore,
\begin{eqnarray}
F = \frac{\partial (n F_n)}{\partial n} \Bigg|_{n\to 1}
~,
\end{eqnarray}
and
\begin{eqnarray} \label{eq:conf_entropy_def}
S_{\rm c} = \beta \frac{\partial F_n}{\partial n} \Bigg|_{n\to 1}
~.
\end{eqnarray}
Performing the functional integration over the field ${\bm \psi}({\bm r})$ we get 
\begin{eqnarray}
F_n = - (\beta n)^{-1} \lim_{g\to 0} \ln \int {\cal D}{\bm m} \int {\cal D}\lambda \exp\Big\{-\frac{\beta}{2} {\tilde {\cal H}}[{\bm m},\lambda] \Big\}
~,
\nonumber\\
\end{eqnarray}
where
\begin{eqnarray} \label{eq:3D_action_replicated}
{\tilde {\cal H}}[{\bm m},\lambda] &=& \sum_{{\bm q},{\bm q}'} \sum_{\alpha,\beta} \sum_{i,j} m^{i,\alpha}_{\bm q} \big[{\tilde G}^{(0)}({\bm q},{\bm q}',\lambda)\big]^{-1}_{ij,\alpha\beta} m^{j,\beta}_{-{\bm q}'}  
\nonumber\\
&-&
\sum_{{\bm q},\alpha} \big( 2{\bm h}^{\alpha}_{-{\bm q}} \cdot {\bm m}^{\alpha}_{\bm q} +\lambda^{(\alpha)}_{\bm q}\big)
~.
\nonumber\\
\end{eqnarray}
Here $\alpha,\beta = 1,\ldots,n$ are replica indices, and we introduced the replicated Green's function
\begin{eqnarray}
\big[{\tilde G}^{(0)}({\bm q},{\bm q}',\lambda)\big]^{-1}_{ij,\alpha\beta} =
\big[G^{(0)}({\bm q},{\bm q}',\lambda)\big]^{-1}_{ij}
\delta_{\alpha\beta} 
+ \frac{g}{n} \delta_{ij}
~.
\nonumber\\
\end{eqnarray}
which is the tensor product of a $3\times 3$ matrix in real space and an $n\times n$ matrix in replica space. The role of the pinning field ${\bm \psi}({\bm r})$ is to break the symmetry of the replicas and to introduce a coupling between them.
After a change of variable (with Jacobian one) the Hamiltonian~(\ref{eq:3D_action_replicated}) is conveniently rewritten as follows:
\begin{eqnarray} \label{eq:3D_action_replicated_changed_variables}
{\tilde {\cal H}}[{\bm m},\lambda] &=& \sum_{{\bm q},{\bm q}'} \sum_{\alpha,\beta} \sum_{i,j} m^{i,\alpha}_{\bm q} \big[{\tilde G}^{(0)}({\bm q},{\bm q}',\lambda)\big]^{-1}_{ij,\alpha\beta} m^{j,\beta}_{-{\bm q}'}  
\nonumber\\
&+&
\sum_{{\bm q},{\bm q}'} \sum_{i,j} h^{i,\alpha}_{\bm q} {\tilde G}^{(0)}_{ij,\alpha\alpha}({\bm q},{\bm q}',\lambda) h^{j,\alpha}_{-{\bm q}'}  
-\sum_{{\bm q},\alpha} \lambda^{(\alpha)}_{\bm q}
~.
\nonumber\\
\end{eqnarray}

We now briefly summarize the calculation. More details are given in what follows. We split $\lambda_{\bm q} = {\bar \lambda}\delta_{{\bm q},{\bm 0}} + {\tilde \lambda}_{\bm q}$, where ${\bar \lambda}$ in the mean-field part of the slave field and ${\tilde \lambda}_{\bm q}$ is its fluctuating part. ${\bar \lambda}$ is determined by requiring the equality $|{\bm m}({\bm r})|^2 = 1$ to be satisfied (on average). The fluctuating part of the slave field introduces higher-order couplings between the components of the magnetization. Assuming the fluctuations to be small, and expanding to second order in ${\tilde \lambda}_{\bm q}$, we obtain an effective fourth-order coupling between the components of the magnetization which is mediated by the propagator $\langle {\tilde \lambda}_{\bm q} {\tilde \lambda}_{-{\bm q}} \rangle$. The latter is nothing but the inverse of the polarizability [see Fig.~\ref{fig:one}b)]. The fourth-order coupling in turn induces self-energy corrections to the full Green's function ${\tilde G}_{ij,\alpha\beta}(q)$. When the number of components of ${\bm m}({\bm r})$ is large, the diagram of Fig.~\ref{fig:one}a) is responsible for the dominant contribution to the self-energy. In what follows we only calculate this contribution. In a self-consistent fashion, the new Green's function is used to re-calculate the polarizability ({\it i.e.} the propagator of the interaction) and the mean field ${\bar \lambda}$. These two quantities are then used to start a new cycle, until convergence is reached. This procedure, named ``self-consistent screening approximation'' (SCSA),~\cite{Bray_prl_1974} is summarized in Fig.~\ref{fig:one}. Finally, the configurational entropy is calculated according to Eq.~(\ref{eq:conf_entropy_def}).

In more detail, we observe that the Hamiltonian~(\ref{eq:3D_action_replicated_changed_variables}) is quadratic in the magnetization field. Therefore, we can integrate out the vector field ${\bm m}({\bm r})$ and obtain an effective action for the slave field $\lambda({\bm r})$. We get
\begin{eqnarray} \label{eq:3D_action_lambda}
{\cal S}[\lambda] &=&- \frac{1}{2} {\rm Tr} \ln \left\{ \big[{\tilde G}^{(0)}({\bm q},{\bm q}',{\bar \lambda})\big]^{-1}_{ij,\alpha\beta} + {\tilde \lambda}^{(\alpha)}_{{\bm q}-{\bm q}'} \delta_{ij} \delta_{\alpha\beta} \right\}  
\nonumber\\
&-&
\frac{\beta}{2} \Bigg[\sum_{{\bm q},{\bm q}'} h^{i,\alpha}_{\bm q}  {\tilde G}^{(0)}_{ij,\alpha\alpha}({\bm q},{\bm q}',\lambda) h^{j,\alpha}_{-{\bm q}'}
-\sum_{{\bm q},\alpha}\lambda^{(\alpha)}_{\bm q}\Bigg]
~.
\nonumber\\
\end{eqnarray}
In Eq.~(\ref{eq:3D_action_lambda}) the trace is on magnetization-direction, momentum and replica indices. The mean-field ${\bar \lambda}$ is determined expanding the right-hand side of Eq.~(\ref{eq:3D_action_lambda}) in powers of ${\tilde \lambda}_{\bm q}$, and setting the first-order term to zero. This procedure leads to the following self-consistent equation~\cite{Chubukov_prb_1994,Starykh_prb_1994,Irkhin_prb_1996}
%It satisfies the following equation
%
\begin{eqnarray} \label{eq:lambdabar_SC}
%\sum_{{\bm q},i} {\tilde G}^{(0)}_{ii}(q) = \frac{1}{T}
%\sum_{{\bm r},i} \frac{\delta^2 \ln (Z)}{\delta h^{i,\alpha}({\bm r})\delta h^{i,\alpha}({\bm r})} = 1
\frac{\delta {\cal S}[\lambda]}{\delta \lambda} \Bigg|_{{\tilde \lambda} = 0} = 0
~,
\end{eqnarray}
%
%Here we took the limit $g\to 0$ and $n\to 1$, and therefore there is no sum over replica indices. 
%Eq.~(\ref{eq:lambdabar_SC}) is generalized, when subleading orders in the $1/N$ expansion are taken into account, to~\cite{Chubukov_prb_1994,Starykh_prb_1994,Irkhin_prb_1996}
which is explicitly rewritten as
\begin{eqnarray} \label{eq:lambdabar_SC_full}
\sum_{{\bm q},i} {\tilde G}^{(0)}_{ii,\alpha\alpha}(q) + \frac{1}{T} \sum_{{\bm q},i,j} h^{i,\alpha}_{{\bm q}} {\tilde G}^{(0)}_{ij,\alpha\alpha}(q) h^{j,\alpha}_{{\bm q}} {\tilde G}^{(0)}_{ji,\alpha\alpha}(q)  = \frac{1}{T}
%~,
\nonumber\\
\end{eqnarray}
where we used that ${\tilde G}^{(0)}_{ij,\alpha\beta}({\bm q},{\bm q}',{\bar \lambda}) \equiv {\tilde G}^{(0)}_{ij,\alpha\beta}(q) \delta_{{\bm q},{\bm q}'}$. No sum over $\alpha$ is understood in Eq.~(\ref{eq:lambdabar_SC_full}), which defines ${\bar \lambda}$ as a function of both the in-plane magnetic field and temperature.

Expanding Eq.~(\ref{eq:3D_action_lambda}) to second order in the fluctuation ${\tilde \lambda}_{\bm q}$ we get
\begin{eqnarray} \label{eq:3D_slave_action_2ord}
{\cal S}[\lambda] &=& \sum_{{\bm q},{\bm q}'} \sum_{i,j,\alpha,\beta} {\tilde \lambda}^{(\alpha)}_{{\bm q}-{\bm q}'} {\tilde G}^{(0)}_{ij,\alpha\beta}(q) {\tilde \lambda}^{(\beta)}_{{\bm q}'-{\bm q}} {\tilde G}^{(0)}_{ji,\beta\alpha}(q')
\nonumber\\
&=&
\sum_{{\bm q}} \sum_{\alpha,\beta} {\tilde \Pi}^{(0)}_{\alpha\beta}({\bm q}) {\tilde \lambda}^{(\alpha)}_{{\bm q}} {\tilde \lambda}^{(\beta)}_{{\bm q}}
~,
\end{eqnarray}
where we introduced the bare polarizability
\begin{eqnarray}
{\tilde \Pi}^{(0)}_{\alpha\beta}({\bm q}) &=& \sum_{{\bm p},i,j} {\tilde G}^{(0)}_{ij,\alpha\beta}(p) {\tilde G}^{(0)}_{ji,\beta\alpha}(|{\bm p} + {\bm q}|)
~.
\end{eqnarray}
The coupling between the magnetization and the slave field induces, to the lowest order in the expansion in ${\tilde \lambda}_{\bm q}$, an effective magnetization-magnetization interaction, whose propagator is $D^{(0)}_{\alpha\beta}({\bm q}) \equiv \big[{\tilde \Pi}^{(0)}({\bm q})\big]^{-1}_{\alpha\beta}$. If the Green's function is dressed by self-energy insertions (subleading in the $1/N$ expansion), the propagator of the interaction is also renormalized, {\it i.e.} $D_{\alpha\beta}({\bm q}) \equiv \big[{\tilde \Pi}({\bm q})\big]^{-1}_{\alpha\beta}$, where~\cite{Chubukov_prb_1994,Starykh_prb_1994,Irkhin_prb_1996}
\begin{eqnarray} \label{eq:full_polarizability}
{\tilde \Pi}_{\alpha\beta}({\bm q}) &=& \sum_{{\bm p},i,j} {\tilde G}_{ij,\alpha\beta}(k) {\tilde G}_{ji,\beta\alpha}(|{\bm p} + {\bm q}|)
~.
\end{eqnarray}
Here ${\tilde G}_{ij,\alpha\beta}(q)$ is the full Green's function. Note that in Eq.~(\ref{eq:full_polarizability}) we have retained only self-energy corrections, and we discarded the vertex ones. This approximation, which greatly simplifies our calculations, is at the heart of the SCSA.~\cite{Bray_prl_1974} Analogously, Eq.~(\ref{eq:lambdabar_SC_full}) is replaced by 
\begin{eqnarray} \label{eq:lambdabar_SC_full2}
\sum_{{\bm q},i} {\tilde G}_{ii,\alpha\alpha}(q) + \frac{1}{T} \sum_{{\bm q},i,j} h^{i,\alpha}_{{\bm q}} {\tilde G}_{ij,\alpha\alpha}(q) h^{j,\alpha}_{{\bm q}} {\tilde G}_{ji,\alpha\alpha}(q)  = \frac{1}{T}
~.
\nonumber\\
\end{eqnarray}
%

%
%Therefore, up to second order in the fluctuations, the action~(\ref{eq:3D_action_replicated}) becomes equivalent to
%%
%\begin{eqnarray} \label{eq:3D_action_replicated_quartic}
%{\tilde S}[{\bm m},\lambda] &=& \sum_{{\bm q},{\bm q}'} \sum_{\alpha,\beta} m^{i,\alpha}_{\bm q} \big[{\tilde G}^{(0)}(q)\big]^{-1}_{ij,\alpha\beta} m^{j,\beta}_{-{\bm q}}  
%\nonumber\\
%&+&
%\frac{1}{2}\sum_{{\bm q}} D^{(0)}(q) \sum_{{\bm k},{\bm k}'} ({\bm m}_{{\bm k}+{\bm q}}\cdot {\bm m}_{{\bm k}}) ({\bm m}_{{\bm k}'-{\bm q}}\cdot {\bm m}_{{\bm k}'})
%~,
%\nonumber\\
%\end{eqnarray}
%%
%where, in the limit $g\to 0$, $D^{(0)}({\bm q}) \equiv 1/{\tilde \Pi}_{\alpha\alpha}({\bm q})$ (no sum over $\alpha$ is understood here --- the diagonal elements are all equivalent).

We assume the full Green's function to have the following one-step replica-symmetry-breaking form:
\begin{eqnarray}
{\tilde G}_{ij,\alpha\beta}(q) = \big[ {\cal G}_{ij}({\bm q}) - {\cal F}_{ij}({\bm q}) \big] \delta_{\alpha\beta} + {\cal F}_{ij}({\bm q})
~,
\end{eqnarray}
where ${\cal F}_{ij}(q)$ is the anomalous (off-diagonal in replica space) part of the Green's function, which is non-zero if the system exhibits broken replica symmetry. Eq.~(\ref{eq:lambdabar_SC_full2}) 
then reads
\begin{eqnarray} \label{eq:lambdabar_SC_fullG}
\sum_{{\bm q},i} {\cal G}_{ii}({\bm q}) + \frac{1}{T} \sum_{{\bm q},i,j} h^{i}_{{\bm q}} {\cal G}_{ij}({\bm q}) h^{j}_{{\bm q}} {\cal G}_{ji}({\bm q}) = \frac{1}{T}
~.
\nonumber\\
\end{eqnarray}
%
%and
%%
%\begin{eqnarray} \label{eq:3D_slave_action_2ord_fullG}
%S[\lambda] &=& \sum_{{\bm q}} \sum_{\alpha,\beta} {\tilde \Pi}_{\alpha\beta}({\bm q}) {\tilde \lambda}^{(\alpha)}_{{\bm q}} {\tilde \lambda}^{(\beta)}_{{\bm q}}
%~,
%\end{eqnarray}
%%
%where now 
The polarizability matrix has the same replica structure, namely ${\tilde \Pi}_{\alpha\beta}(q) = \big[ \Pi_{\cal G}(q) - \Pi_{\cal F}(q) \big] \delta_{\alpha\beta} + \Pi_{\cal F}(q)$, where
\begin{eqnarray} \label{eq:3D_Pi_def}
&& \Pi_{\cal G}({\bm q}) = \sum_{{\bm p},i,j} {\cal G}_{ij}({\bm p}) {\cal G}_{ji}({\bm p}+{\bm q})
~,
\nonumber\\
&& \Pi_{\cal F}({\bm q}) = \sum_{{\bm p},i,j} {\cal F}_{ij}({\bm p}) {\cal F}_{ji}({\bm p}+{\bm q})
~.
\end{eqnarray}
The propagator of the magnetization-magnetization interaction is ${\tilde D}_{\alpha\beta}({\bm q}) \equiv \big[ {\tilde \Pi}({\bm q}) \big]^{-1}_{\alpha\beta} = \big[ D_{\cal G}(q) - D_{\cal F}(q) \big] \delta_{\alpha\beta} + D_{\cal F}(q)$ where, in the limit $n\to 1$,
\begin{eqnarray} \label{eq:app_interaction}
D_{\cal G}({\bm q}) &=& \Pi_{\cal G}^{-1}({\bm q})
~,
\nonumber\\
D_{\cal F}({\bm q}) &=&  -\frac{D_{\cal G}^2({\bm q}) \Pi_{\cal F}({\bm q})}{1 - D_{\cal G}({\bm q}) \Pi_{\cal F}({\bm q})}
~.
\end{eqnarray}
The interaction $D_{\cal G}({\bm q})$, which extends in principle up to the upper momentum $\Lambda$, needs to be cutoff at momenta of the order of $2 q_0$ for our theory to produce meaningful results. The qualitative behavior of $S_{\rm c}$ is only weakly affected by this truncation procedure but we find that, if no cut-off is introduced on the interaction, the self-consistent procedure does not converge and the self-energy strongly depends on $\Lambda$.
Finally, the self-energy is ${\tilde \Sigma}_{ij,\alpha\beta}(q) = \big[ \Sigma_{{\cal G},ij}(q) - \Sigma_{{\cal F},ij} (q) \big] \delta_{\alpha\beta} + \Sigma_{{\cal F},ij}(q)$, where
\begin{eqnarray} \label{eq:3D_Sigma_def}
&& \Sigma_{{\cal G},ij}({\bm q}) = 2\sum_{\bm p} D_{\cal G}({\bm p}) {\cal G}_{ij}({\bm p}+{\bm q})
~,
\nonumber\\
&& \Sigma_{{\cal F},ij}({\bm q}) = 2\sum_{\bm p} D_{\cal F}({\bm p}) {\cal F}_{ij}({\bm p}+{\bm q})
~.
\end{eqnarray}
Since ${\tilde G}^{-1}_{ij,\alpha\beta} (q) = \big[G^{(0)}(q)\big]^{-1}_{ij,\alpha\beta} + {\tilde \Sigma}_{ij,\alpha\beta}(q)$, when $n\to 1$
\begin{eqnarray}
{\cal G}({\bm q}) &=& \Big\{ \big[G^{(0)}({\bm q})\big]^{-1} + {\bar \lambda} \openone + \Sigma_{\cal G}({\bm q})\Big\}_{ij}^{-1}
~,
\nonumber\\
{\cal F}({\bm q}) &=&  -{\cal G}({\bm q}) \Sigma_{\cal F}({\bm q}) {\cal G}({\bm q}) \big[1 - \Sigma_{\cal F}({\bm q}) {\cal G}({\bm q}) \big]^{-1}
~,
\end{eqnarray}
where the matrix product on the spin indices is understood.

The configurational entropy is determined from the derivative of the free energy~\cite{Luttinger_pr_1960}
\begin{eqnarray}
\frac{F_n}{T} = \frac{1}{2n}\Big({\rm Tr}\ln{\tilde G}^{-1} + {\rm Tr}({\tilde \Sigma} {\tilde G}) - \Phi[{\tilde G}] - \frac{{\rm Tr} ({\bm h}{\tilde G}{\bm h})}{2T} \Big)
\nonumber\\
\end{eqnarray}
with respect to $n$. Here the trace is over momentum, spin and replica indices, while the Luttinger-Ward functional~\cite{Luttinger_pr_1960} $\Phi[{\tilde G}]$ is determined by the choice of the self-energy and by the fact that
\begin{eqnarray} \label{eq:functional_derivative}
\frac{\delta\Phi[{\tilde G}]}{\delta {\tilde G}_{ij,\alpha\beta}({\bm q})} = {\tilde \Sigma}_{ij,\alpha\beta}({\bm q})
~,
\end{eqnarray}
which leads to $\Phi[{\tilde G}] = {\rm Tr}\ln{\cal D}^{-1}$. The entropy is then given by~\cite{Schmalian_prl_2000}
\begin{eqnarray} \label{eq:SF_S_c_final}
\frac{S_{\rm c}(T)}{V} &=& - \frac{1}{2} \sum_{\bm q} \Bigg\{ {\rm Tr}\Big( \ln\big[ 1 - {\cal G}^{-1}({\bm q}) {\cal F}({\bm q}) \big] 
\nonumber\\
&+&
{\cal G}^{-1}({\bm q}) {\cal F}({\bm q}) \Big)
-\ln\left[ 1 - \frac{\Pi_{\cal F}({\bm q})}{\Pi_{\cal G}({\bm q})} \right] - \frac{\Pi_{\cal F}({\bm q})}{\Pi_{\cal G}({\bm q})} \Bigg\}
\nonumber\\
&-&
\frac{1}{4T} \sum_{{\bm q},i,j} h^{i}_{-{\bm q}} {\cal F}_{ij}({\bm q}) h^{j}_{{\bm q}}
~.
\nonumber\\
\end{eqnarray}
Here the trace is on the three magnetization directions. Note that the term in the last line involves the anomalous Green's function in the direction of the external magnetic field. If the latter is parallel to the plane of the 2D system the configurational entropy does not depend explicitly of ${\bm h}_{\bm q}$, since the in-plane components of the anomalous Green's function vanish.

As in the conventional theory of phase transitions,~\cite{Schieffer_Theory_Superconductivity} the coupling $g\to 0^+$ can lead to the breaking of the replica symmetry. Such a scenario is encoded in the self-energy $\Sigma_{\alpha\beta}({\bm q})$. The starting point of the self-consistent calculation outlined in Fig.~\ref{fig:one} is a Green's function which has small off-diagonal components in replica space. At the end of the self-consistent loop, the self-energy can have finite off-diagonal components. This is however not sufficient to say that the system is a stripe glass. It is necessary to calculate the configurational entropy according to Eq.~(\ref{eq:SF_S_c_final}) and, if $S_{\rm c}>0$, we say that for the chosen parameters (temperature and magnetic field) the system is in the glass phase. Conversely, a negative $S_{\rm c}$ shows that our theory breaks down and cannot properly describe the glassy state. It would probably be necessary to consider the full-replica-symmetry-breaking solution to get the correct value of the configurational entropy, but this task is beyond the scope of the present work. Here we focus on the transition between the ordered stripe phase and the glassy state. We identify $T_{\rm K}$, {\it i.e.} the temperature at which the configuration entropy vanishes linearly, as the transition temperature.

\section{Results}
\label{sect:results}
To perform the numerical calculations we rescale all energies in units of $J$, and all lengths in units of the inverse of the upper cutoff of the momentum integration ($\Lambda$). Typical values of the exchange coupling are given in Refs.~\onlinecite{Brucas_prb_2004,Brucas_prb_2008,Hafermann_apl_2009} for the case of ${\rm Fe}_{81} {\rm Ni}_{19}/{\rm Co}~(001)$ multilayers. Once converted to two-dimensional units ({\it i.e.} using that the height of a bilayer is $\sim 4.2~{\rm nm}$), we get $J\sim 0.1~{\rm eV}$. $\Lambda^{-1}$ is instead of the order of the thickness of domain walls [$\ell_{\rm DW}\sim \sqrt{J/K} \sim 20~{\rm nm}$ for ${\rm Fe}_{81} {\rm Ni}_{19}/{\rm Co}~(001)$ multilayers~\cite{Brucas_prb_2004,Brucas_prb_2008,Hafermann_apl_2009}]. Approximating domain walls with dimensionless lines we implicitly coarse-grain our system in cells of lateral size $\sim \ell_{\rm DW}$. Therefore, its properties depend on the ratio between such scale and the pattern wavelength.~\cite{Bray_advphys_1994,Bray_advphys_2002}
%An estimate for the anisotropy $K \Lambda^{-2}$ is also given,~\cite{Brucas_prb_2004,Brucas_prb_2008,Hafermann_apl_2009} and it turns out to be of the order of $J$.~\cite{Brucas_prb_2004,Brucas_prb_2008,Hafermann_apl_2009} The value of the cutoff $\Lambda$ is determined in what follows.

\begin{figure}[t]
\begin{center}
\begin{tabular}{c}
\includegraphics[width=0.99\columnwidth]{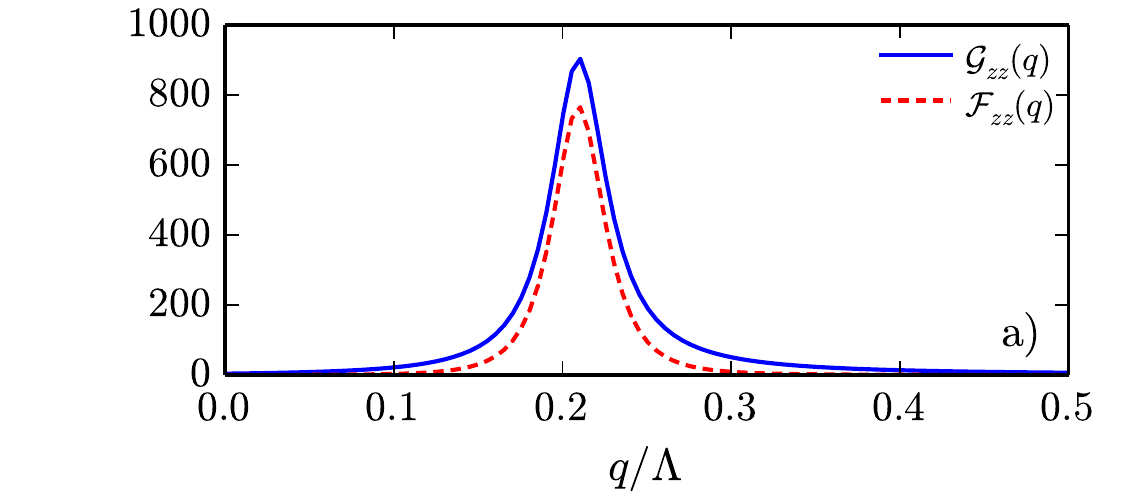}
\\
\includegraphics[width=0.99\columnwidth]{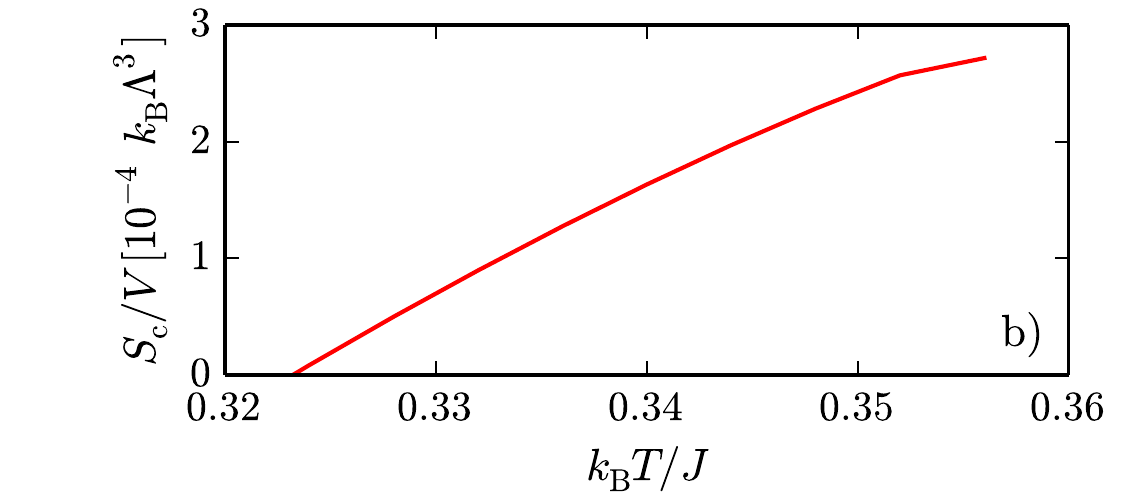}
\end{tabular}
\end{center}
\caption{
Panel a) the Green's functions ${\cal G}_{zz}(q)$ and ${\cal F}_{zz}(q)$ at the Kauzmann temperature $T_{\rm K}$ in units of $1/J$, plot as a function of the momentum $q$ (in units of the upper momentum cutoff $\Lambda$). In this plot we set $h=0$, $K/(J \Lambda^2)=0.2$, $d \Lambda=20$, and $Q/(J \Lambda^3)=0.02$. Both the functions are peaked at $q_0\sim 0.2~\Lambda$, which is the modulus of the wavevector of the striped phase. 
%${\cal F}_{zz}(q)$ decays faster than ${\cal G}_{zz}(q)$ when $q$ is moved away from $q_0$, since defect wandering destroys long-time correlations of a non-perfect striped phase much faster than the instantaneous ones. 
%
Panel b) the configurational entropy for a 2D Heisenberg ferromagnet subject to a long-range dipole-dipole interaction.
%We rescaled all the energies in units of the exchange coupling $J$, and all the lengths in units of the inverse of the upper momentum cutoff $\Lambda^{-1}$. 
The parameters of this plot are the same as in panel a).
\label{fig:two}}
\end{figure}

In Fig.~\ref{fig:two}a) we show the form factor of the magnetization along the ${\hat {\bm z}}$ direction ${\cal G}_{zz}(q) = \langle m^{z}_{{\bm q}} m^{z}_{-{\bm q}} \rangle$. In this plot we set $h=0$, $K/(J\Lambda^2)=0.2$, $d\Lambda=20$, and $Q/(J\Lambda^3)=0.02$. This choice of $K$ implies that $\Lambda^{-1} \lesssim \ell_{\rm DW}$.
%We find indeed that, to have a maximum of ${\cal G}_{zz}(q)$ at finite $q$, we need the anisotropy $K$ to be of the order of $J\Lambda^2$. The value of the cutoff $\Lambda$ is determined in what follows.
In our calculation we set $\Sigma_{\cal G}({\bm q}) = 0$. This implies that ${\cal G}_{ij}$ is diagonal in the spin indices and it is equal to its non-interacting counterpart. This approximation is justified by the fact that $\Sigma_{\cal G}({\bm q})$ is very small and only weakly dependent on ${\bm q}$. As a consequence, for a sufficiently large value of $K/(J\Lambda^2)$, ${\cal G}_{ii}(q)$ is maximum at $q = 0$ for $i=x,y$, while ${\cal G}_{zz}(q)$ is peaked at $q=q_{\rm 0} \neq 0$. For our choice of the parameters, $q_0 \sim 0.2~\Lambda$. The finite value of $q_0$ makes it possible to form a low-temperature striped phase. However, since the Hamiltonian is rotationally invariant in momentum space, a stripe glass can emerge at sufficiently high temperature.~\cite{Schmalian_prl_2000} Note that no glass (or striped phase) emerges if ${\cal G}_{zz}(q)$ is peaked at $q=0$, {\it i.e.} when $K/(J\Lambda^2)$ is too small. 
%Therefore, the strong out-of-plane anisotropy turns out to beneficial for the formation of both the ordered and glassy phases.
The period of the low-temperature stripe phase $L_{\rm th} =2\pi/q_0$ should be compared with the experimentally measured one~\cite{Brucas_prb_2004} $L_{\rm exp} \simeq 460~{\rm nm}$. This leads to $\Lambda^{-1} \sim 14~{\rm nm}$.

The magnitude of the magnetization can also be extracted from Ref.~\onlinecite{Brucas_prb_2004}, where it is said to be of the order of $\sim 2 \mu_{\rm B}/{\rm atom}$. Here $\mu_{\rm B} \simeq 5.79 \times 10^{-5}~{\rm eV}/{\rm T}$ is the Bohr magneton. Since the lattice constant of the samples of Ref.~\onlinecite{Brucas_prb_2004} is $a_0 \simeq 0.282~{\rm nm}$, every square of side $\Lambda^{-1}$ contains about $2,700$ atoms. Therefore, $M\simeq 0.3~{\rm eV}/{\rm T}$. For future reference, we recall that $h = H_\parallel M \Lambda^{2}/2$, where $H_\parallel$ is the physical in-plane magnetic field (measured in units of Tesla).

In Fig.~\ref{fig:two}b) we show the configurational entropy for a 2D Heisenberg ferromagnet subject to a long-range dipole-dipole interaction. The parameters are the same as in Fig.~\ref{fig:two}a). At large temperature, the system is in a liquid (or paramagnetic) state, in which the magnetization is completely randomized. The anomalous part of the Green's function ${\cal F}_{ij}({\bm q})$ vanishes, and so does the configurational entropy. At the temperature $T_{\rm A}$, $S_{\rm c}$ jumps from zero to a finite positive value. Below the temperature $T_{\rm A}$ the system is found in a glassy state. The configurational entropy decreases with decreasing temperature, and vanishes at the Kauzmann temperature $T_{\rm K}<T_{\rm A}$. Below this temperature, the theory predicts an unphysically negative $S_{\rm c}$. Such a behavior shows that the {\it Ansatz} for the Green's function is not correct for $T<T_{\rm K}$.
Below $T_{\rm K}$ the system is believed to be able to still form a glass but with an extremely long time scale.~\cite{Schmalian_prl_2000,Westfahl_prb_2001,Westfahl_prb_2003,Wu_prb_2004} Such a regime is not only beyond the scope of our work, but we also believe that, as observed experimentally,~\cite{Brucas_prb_2004,Brucas_prb_2008,Hafermann_apl_2009} below $T_{\rm K}$ the system forms a striped phase by spontaneously breaking the rotational symmetry.

%%%%%%%%%%%%%%%%%%%
%%%%%%%%%%%%%%%%%%%

%In our calculation we set $\Sigma_{\cal G} = 0$. This implies that ${\cal G}_{ij}$ is diagonal in the spin indices, and is equal to its non-interacting counterpart $G^{(0)}_{ij}(q)$. Therefore, ${\cal G}_{ii}$ is maximum at $q = 0$ for $i=x,y$, while ${\cal G}_{zz}(q)$ is peaked at $q=q_{\rm 0} \neq 0$. The value of $q_0$ depends on the parameters of the problem, and it is non-zero for a sufficiently large value of the anisotropy $K$. The finite value of $q_0$ is responsible for the existence of the striped-domain phase, whose modulation wavevector has modulus $q_0$ and orientation due to the mechanism of spontaneously broken rotational symmetry. Since all directions are in principle equivalent, a stripe glass can emerge in a certain range of temperatures.~\cite{Schmalian_prl_2000} It is clear that no glass (or striped phase) can emerge if the peak of ${\cal G}_{zz}(q)$ occurs for $q=0$, as it happens for too weak out-of-plane anisotropies. A strong out-of-plane anisotropy is therefore beneficial for the formation of a striped glass.

The anomalous Green's function ${\cal F}_{ii}(q)=0$ for $i=x,y$, while ${\cal F}_{zz}(q)$ is non-zero and peaked at $q=q_0$ [see Fig.~\ref{fig:two}a)]. We recall~\cite{Schmalian_prl_2000,Westfahl_prb_2001} that ${\cal G}_{ij}(q)$ [${\cal F}_{ij}(q)$] contains the information about the short- [long-] time correlations of the system. Both functions are maximized by the perfect stripe arrangement, and the widths of their peaks define two length scale, $\xi_{\cal G}$ and $\xi_{\cal F}$.~\cite{Westfahl_prb_2001} The former is interpreted as the correlation function of the stripe phase, which diverges at zero temperature, while the latter is the length scale over which defects can wander. Note that ${\cal F}_{zz}(q)$ decays faster than ${\cal G}_{zz}(q)$, when $q$ is moved away from $q_0$. Long-time correlations of non-perfect stripe phases are suppressed with respect to the instantaneous ones since defects can wonder and destroy them.~\cite{Westfahl_prb_2001} At the Kauzmann temperature, $\xi_{\cal G} \sim 7~\mu{\rm m}$ and $\xi_{\cal F} \sim 2~\mu{\rm m}$. Samples of lateral size smaller than $\xi_{\cal G}$ are therefore expected to exhibit an almost striped phase, with an average inter-defect distance of the order of $\xi_{\cal F}$. Indeed, two defects that are at the distance smaller than $\xi_{\cal F}$ can wander and annihilate each other. 
%The length $\xi_{\cal F}$ increases with the temperature, and reaches $\xi_{\rm F}\sim 2.8~\mu{\rm m}$ at $T=T_{\rm A}$. 

\begin{figure}[t]
\begin{center}
\begin{tabular}{c}
\includegraphics[width=0.99\columnwidth]{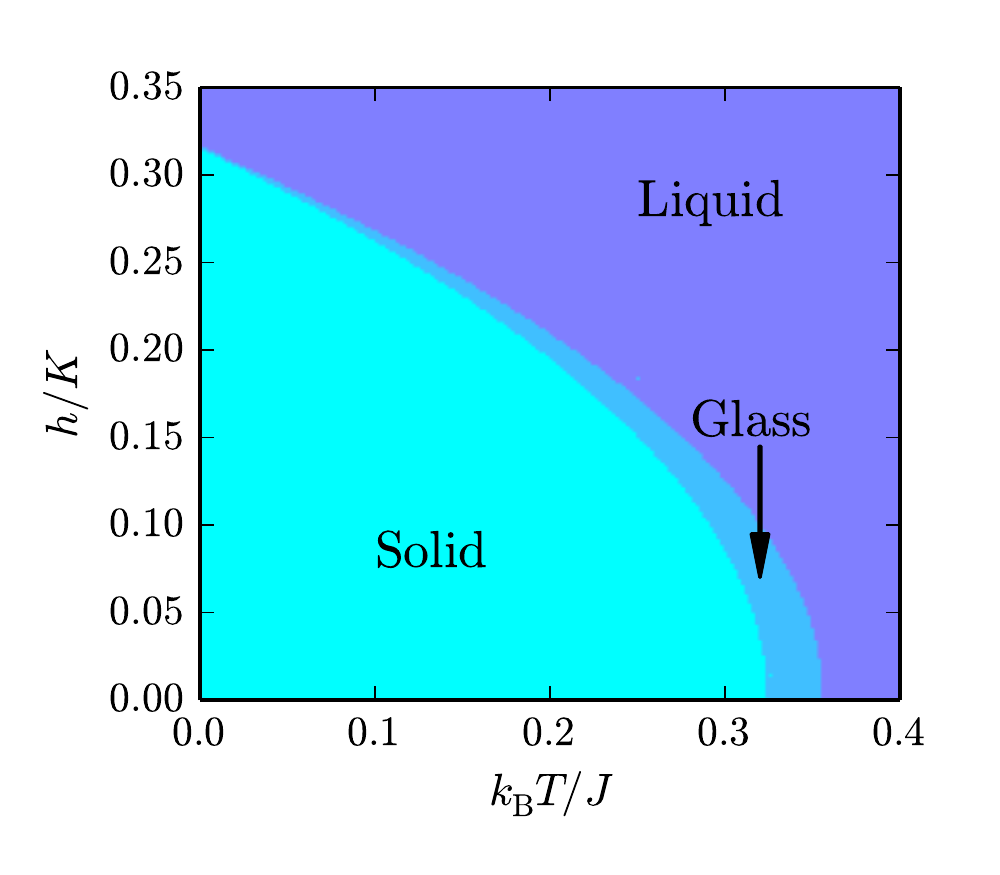}
\end{tabular}
\end{center}
\caption{
The phase diagram of the magnetic system described by the action~(\ref{eq:3D_action_def}). Note that the temperature of the glass transition shifts to lower temperatures as the in-plane magnetic field is increased. 
%This in turn implies that, in an hypothetical experiment in which the temperature of the layered magnetic system is kept fixed, the magnetic field can be used to tune it in a glassy phase.~\cite{Brucas_prb_2004}
\label{fig:three}}
\end{figure}

It is interesting to study the effect of a uniform in-plane magnetic field, for example along the ${\hat {\bm y}}$ direction, {\it i.e.} ${\bm h}_{\bm q} = \delta_{{\bm q},0}h {\hat {\bm y}}$, which directly contributes to the configurational entropy though the term on the last line of Eq.~(\ref{eq:SF_S_c_final}).
%In the present theory, its effect is twofold. On one hand, Eq.~(\ref{eq:main_lambdabar_SC}) implies that the magnetic field affects the value of the mean field ${\bar \lambda}$. On the other hand, it contributes directly to the configurational entropy though the last term on the right-hand side of Eq.~(\ref{eq:main_F_n_luttinger}). 
Since ${\cal F}_{yy}=0$, there is no direct effect of the magnetic field on the configurational entropy. This does not mean that there is no effect at all. The phase diagram of the system (Fig.~\ref{fig:three}) is completely determined by the value of ${\bar \lambda}$, which depends on both the temperature and magnetic field via Eq.~(\ref{eq:lambdabar_SC_fullG}).

%%%%%%%%%%%%%%%%%%%
%%%%%%%%%%%%%%%%%%%

Note that the stripe-glass phase shifts to lower temperatures when $h$ is increased. Therefore, starting from the ordered striped phase and by applying an in-plane magnetic field, defects can be introduced into the system and a glass can be formed. It is interesting to note that a similar behavior has been observed experimentally~\cite{Brucas_prb_2004,Seul_prl_1992} for the case of ${\rm Fe}_{81} {\rm Ni}_{19}/{\rm Co}~(001)$ multilayers, in which the application of an in-plane magnetic field induces induces a chaotization of the domain-wall structure. Such a behavior is compatible with the self-induced glassiness discussed in this Paper. The value of the critical magnetic field to enter the glass phase at a temperature $T/J=0.1$ (which roughly corresponds to room temperature, for $J=0.1~{\rm eV}$) is $h/K\sim 0.25$. This corresponds to a magnetic field $H_\parallel \sim 30~{\rm mT}$, whose order of magnitude is in good agreement with what reported in Ref.~\onlinecite{Brucas_prb_2004}.

\section{Conclusions}
\label{sect:conclusions}
In this paper we have discussed the emergence of self-induced glassiness in two-dimensional magnetic thin-film multilayers. These systems have attracted a great deal of interests in the recent years because of the intriguing behavior of the domain-wall structure.~\cite{Seul_science_1995,Donzelli_jap_2003,Hellwig_physB_2003,Favieres_jap_2002,Onoue_jap_2002,Belliard_jap_1997,Barnes_jap_1994,Hamada_jmmm_1999,Ausanio_jmmm_2001,Zeper_jap_1989,Louail_jmmm_1997,Ploessl_jap_1993,Tonomura_prb_1982,Scheinfein_prb_1991,Schafer_IEEE_1992,Eimuller_jap_2000,Castrucci_prb_2002,Asenjo_prb_2000,Hamada_jmmm_2002,Ausanio_jmmm_2002,Asenjo_jmmm_1999,Brucas_prb_2004,Brucas_prb_2008,Hafermann_apl_2009,Prudkovskii_epl_2006}
 Due to the many possible applications in the everyday life, it is fundamental to improve our understanding of their properties. 

At low temperature, the out-of-plane anisotropy and the competition between the short-range ferromagnetic coupling and long-range dipole-dipole interactions leads to the formation of an ordered phase.~\cite{Choe_prb_1999,Ng_prb_1995,Whitehead_jpcm_1994,Kittel_pr_1946,Garel_prb_1982,Yafet_prb_1988,Kaplan_jmmm_1993} The spins are mainly oriented out of the film plane and their direction switches periodically from up to down, thus forming a stripe domain pattern. The modulation wavelength of the pattern is due to the different scales of interactions, while the stripe direction emerges as a result of the spontaneous breaking of the rotational symmetry.

The equivalence of the in-plane directions is responsible for the formation of a stripe glass.~\cite{Schmalian_prl_2000,Westfahl_prb_2001,Westfahl_prb_2003,Wu_prb_2004} As the temperature increases, the striped phase starts to melt and defects to wander.~\cite{Schmalian_prl_2000,Westfahl_prb_2001} Eventually, a glass is formed. This phase is characterized by the appearance of many metastable states, whose number compensates for their small statistical weight. The system is then trapped in one of the minima for a long time, and its behavior becomes  non-ergodic. This is reflected in the finite configurational entropy ($S_{\rm c}$).

The glass is formed in a quite narrow range of temperatures, between the Kauzmann temperature~\cite{Kauzmann_chemrev_1948} $T_{\rm K}$ and the melting temperature~\cite{Schmalian_prl_2000} $T_{\rm A}$, above which the system is found in a paramagnetic state. At $T_{\rm A}$ the configurational entropy jumps from zero to a finite value,~\cite{Schmalian_prl_2000,Westfahl_prb_2001,Westfahl_prb_2003,Wu_prb_2004} therefore signaling a first-order glass transition. The entropy then vanishes linearly, as the temperature is decreased, at $T_{\rm K}$. Below $T_{\rm K}$ the system is in the ordered phase.

The application of an in-plane magnetic field shifts the glassy phase towards lower temperatures. Interestingly, experiments~\cite{Brucas_prb_2004} have observed a chaotization of the domain-wall pattern in layered structures when an in-plane magnetic field is applied. Our theory is compatible with the observed behavior: the system forms a glass when $T_{\rm K}$ becomes equal to the experimental temperature. At room temperature, the magnetic field that leads to the transition is $H_\parallel \sim 30~{\rm mT}$, which is in good agreement with what reported in experiments.~\cite{Brucas_prb_2004}

\section{Acknowledgements}
This work was supported by the Nederlandse Wetenschappelijk Organisatie (NWO) via the Spinoza Prize.

\end{document}